\documentclass[sigconf]{acmart}

\usepackage{multirow}
\usepackage{url}
\usepackage{xcolor}

\newcommand{\figref}[1]{Figure \ref{#1}}
\newcommand{\tabref}[1]{Table \ref{#1}}
\newcommand{\modelname}{PiCoGen}
\newcommand{\tokencls}[1]{{#1}}
\newcommand{\token}[1]{\texttt{[#1]}}

\AtBeginDocument{%
  }

\copyrightyear{2024}
\acmYear{2024}
\setcopyright{acmlicensed}\acmConference[ICMR '24]{Proceedings of the 2024 International Conference on Multimedia Retrieval}{June 10--14, 2024}{Phuket, Thailand}
\acmBooktitle{Proceedings of the 2024 International Conference on Multimedia Retrieval (ICMR '24), June 10--14, 2024, Phuket, Thailand}
\acmDOI{10.1145/3652583.3657626}
\acmISBN{979-8-4007-0619-6/24/06}

\begin{document}

\title{PiCoGen: Generate Piano Covers with a Two-stage Approach}


\author{Chih-Pin Tan}
\affiliation{%
  \institution{National Taiwan University, KKCompany Technologies}
  \streetaddress{No. 1, Sec. 4, Roosevelt Rd.}
  \city{Taipei}
  \country{Taiwan}
  }
\email{tanchihpin0517@gmail.com}

\author{Shuen-Huei Guan}
\affiliation{%
  \institution{KKCompany Technologies}
  \streetaddress{}
  \city{Taipei}
  \country{Taiwan}}
\email{drakeguan@kkcompany.com}

\author{Yi-Hsuan Yang}
\affiliation{%
  \institution{National Taiwan University}
  \streetaddress{No. 1, Sec. 4, Roosevelt Rd.}
  \city{Taipei}
  \country{Taiwan}}
\email{yhyangtw@ntu.edu.tw}

\begin{abstract}
Cover song generation 
stands out as 
a popular way of music making
in the music-creative community. In this study, we introduce Piano Cover Generation (PiCoGen),  a two-stage approach for automatic cover song generation that transcribes the melody line and chord progression of a song given its audio recording, and then uses the resulting lead sheet as the condition to generate a piano cover in the symbolic domain. 
This approach is advantageous in that it does not required paired data of covers and their original songs for training. 
Compared to an existing approach that demands such paired data, our evaluation shows that PiCoGen demonstrates competitive or even superior performance across songs of different musical genres.
\end{abstract}




\maketitle

\section{Introduction}

Cover song generation, recreating or rearranging the musical elements from an existing piece, is popular within the music-creative community.
Musicians craft a cover song with careful consideration of musical components such as melody, chords, rhythm, and performance techniques, in a style or musical genre that may be different from the original piece (e.g., Pop to Jazz), and with different sets of instruments (e.g., using only the piano or the guitar).

Attempts have also been made to create cover songs automatically,
relying on a supervised approach trained on paired data of songs and their human-made cover versions.
Takamori \emph{et al.} \cite{takamori2019audio} employ a regression model with acoustic features and structural analysis to generate 
piano covers.
Song2Guitar \cite{ariga2017song2guitar} uses a hidden Markov model 
to generate guitar covers with fundamental frequency (f0), beat, and chord information.
More recently, Choi \emph{et al.} present the Pop2Piano model \cite{choi2023pop2piano}, which represents the first attempt that uses 
deep learning 
for cover song generation.
Specifically, it 
leverages the MT3 architecture \cite{gardner2021mt3}, an encoder-decoder Transformer designed to convert an audio signal into a sequence of discrete tokens, to create piano covers 
of music.
In collecting paired data needed for model training, 
Choi \emph{et al.} find it important to employ heuristics to improve the data quality,
involving a matching routine to identify covers of songs from the Internet,
and an alignment routine to 
synchronize the songs and their covers. 

\begin{figure}
  \centering
  \includegraphics[width=\linewidth]{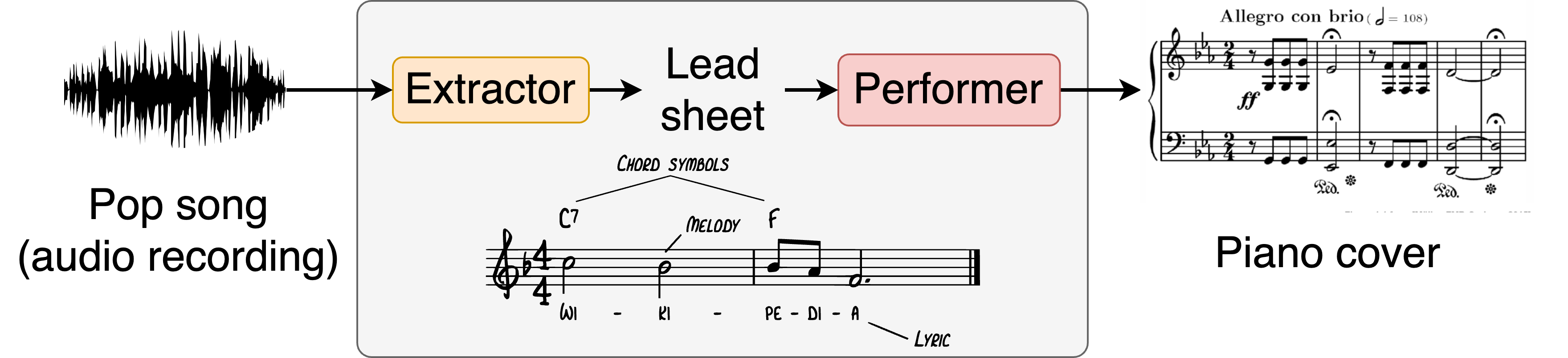}
  \caption{An overview of PiCoGen. The model generates a piano cover in two stages: extracts firstly a lead sheet (i.e., melody line and chord progression) from an audio recording of the original song via  audio analysis (i.e., transcription),
  and then turns the extracted lead sheet into a piano performance via conditional symbolic-domain music generation.
  }
  \label{fig:overview}
\end{figure}

To get rid of the reliance on paired data and to supply deep learning models with more training data,
we explore instead in this paper a novel approach for piano cover generation that does not need paired data at all. Our approach draws inspiration from musical domain knowledge that a cover and its original song tend to share a common melody line and chord progression, and therefore uses the symbolic-domain \emph{lead sheet}~\cite{liu2018lead,wu2023compembellish} (i.e., a MIDI-like representation of melody$+$chord)\footnote{The lead sheet is strongly linked to the f0 contours and chroma features of music, both of which play significant roles in cover song identification \cite{serra2010audio, yesiler2019tacos, silva2018summarizing}.}
as the intermediate gateway transferring a musical audio involving arbitrary instrumentation into a piano-only rendition. Specifically, as depicted in Figure \ref{fig:overview}, we propose a two-stage model that firstly extracts the lead sheet directly from the song audio, and then generates a piano cover given the extracted lead sheet. We refer to the resulting model as
\textbf{Pi}ano \textbf{Co}ver \textbf{Gen}eration (PiCoGen).

The previous Pop2Piano model  \cite{choi2023pop2piano} uses a \emph{piano transcription}-like approach and creates piano covers with a \emph{single} stage of audio-to-symbolic conversion 
with no explainable intermediate representation, trained using paired data of original music and  piano covers.  
In contrast, PiCoGen bypasses the need of such a paired data by decomposing the conversion task into \emph{two} stages.
PiCoGen needs other two types of paired data, one for each stage, yet both of which have been available in the research community through efforts on audio-to-symbolic lead sheet transcription \cite{matti2008auto, weil2009automatic, donahue2022melody} and symbolic-to-symbolic conditional piano  generation \cite{hsiao2021compound, wu2023compembellish}.


As its name implies, Pop2Piano is evaluated on Pop music only~\cite{choi2023pop2piano}. However, we set forth to investigate how well PiCoGen 
compares to Pop2Piano on generating the piano covers for up to ten different musical genres, including cases where lead sheet transcription can be difficult. This is to study to which extent the domain knowledge (or assumptions) incorporated by PiCoGen is adequate, and to evaluate the generalizability of both Pop2Piano and PiCoGen.

We invite readers to our project website\footnote{\url{https://tanchihpin0517.github.io/PiCoGen}} for audio examples of the generated piano covers and the source code of this project.

\section{Background}

Piano cover generation is a special class of cover song generation where the target output is a piano-only rearrangement of a given music piece. It is expected that the generated cover and the original piece can be identified as different versions of the same music composition, and that the cover itself is pleasant to listen to.

Due to the availability of high-quality piano synthesizers, piano cover generation can be approached via generating piano music in a symbolic representation such as Musical Instrument Digital Interface (MIDI), instead of generating audio signals directly. This is the approach taken by Pop2Piano and similarly this work. As such, piano cover generation is related to the following three tasks.

\emph{Automatic Music Transcription} (AMT) involves converting music signals into symbolic representations such as 
piano rolls.
\emph{Automatic piano transcription}, for example, aims to generate a musical score-like transcription of all the notes involved in an audio recording of piano performance \cite{benetos2019amt,toyama2023apt, hawthorne2017onsets, hawthorne2021sequence, gardner2021mt3, kong2021high}. \emph{Lead sheet transcription}, as another instance of AMT, transcribes the notes corresponding to the melody line and recognizes the chord names involved in the harmonic progression 
\cite{matti2008auto, weil2009automatic, donahue2022melody}.
Pop2Piano adopts MT3 \cite{gardner2021mt3}, the state-of-the-art of piano transcription, as their model backbone as both piano transcription and piano cover generation convert audio into piano scores. A main difference here is that MT3 assumes its input to be audio recordings of piano-only music, so Pop2Piano needs paired data of \{original songs, piano covers\} for supervised training to cope with input with arbitrary instrumentation.


\emph{Symbolic-domain music generation} aims to generate novel pieces of music in a symbolic format, usually by representing a music piece as a sequence of discrete tokens 
\cite{huang2018music, huang2020pop, hsiao2021compound, wu2023compembellish,von2022figaro,dong2023multitrack, lu2023musecoco}. The generation process can be either \emph{unconditional} (i.e., from-scratch generation) or \emph{conditional}. A key observation of our work is that piano cover generation can be viewed as a conditional generation task---generating token sequences of piano performances conditioned on pre-given lead sheets. Such a symbolic-to-symbolic conditional piano generation problem has been tackled before \cite{hsiao2021compound, wu2023compembellish}. 


\emph{Music style transfer} involves the process of altering the style of a music piece to match the style of a provided example, influenced by various attributes such as orchestration, chord progression, and tonality \cite{roberts2018hierarchical,cifka2020groove2groove,wu2021musemorphose}. 
Cover song generation can be considered as a specific type of music style transfer in general.

\section{\modelname}
Given an audio segment $A$ of a song,
cover song generation establishes a model $f: A \rightarrow S$ that creates an alternative version of the input. For piano cover generation, the output $S$ is pure piano music, which can be represented in the so-called symbolic domain with discrete tokens bearing explicit musical meaning \cite{oore2018time}.
While many token representations for symbolic music have been proposed in the literature \cite{miditok2021},
Pop2Piano adopts the \emph{MIDI-like} representation \cite{huang2018music}, which uses tokens that indicate the pitch, onset time, offset time, and velocity (which is related to perceptual loudness) of each  note involved in a piano playing. Pop2Piano can therefore be viewed as a cross-domain sequence-to-sequence model converting a segment of continuous audio waveform to a sequence of discrete tokens.

PiCoGen is different from Pop2Piano mainly in two aspects: the model architecture (two-stage vs. single-stage), and the adopted token representation for symbolic music.
We provide details below.



\subsection{Proposed Two-stage Model}

PiCoGen decomposes $f$ into two steps, $f=e \circ g$. 
The first step $e: A\rightarrow L$ converts the audio input into an intermediate representation $L$, while the second step $g: L \rightarrow S$ generates a cover based only on $L$.
In other words, this two-stage approach assumes $A$ and $S$ are conditional independent given $L$.
As shown in Figure \ref{fig:overview}, a fundamental assumption of PiCoGen is that the lead sheet can serve as such an intermediate representation.
When $A$ and $S$ share the same underling lead sheet, we assume that they would sound like the same song.
From a style transfer viewpoint, lead sheet is regarded as the ``content'' to be reserved.
Unlike style transfer in general, the content here is a human-readable lead sheet, facilitating interpretable and controllable music generation.

As shown in \figref{fig:overview},
\modelname{} accordingly contains two sub-models, the \emph{Extractor} that performs $e$ and outputs the corresponding lead sheet sequence $L$ of the input audio $A$, and the \emph{Performer} that performs $g$ and generates piano token sequence $S$ based on the transcribed lead sheet.
Here $e$ and $g$ can be viewed as two sequence-to-sequence problems. In both cases, the objectives are to maximize the probability of the target sequence given the source sequence.

\begin{figure}
  \centering
  \includegraphics[width=0.9\linewidth]{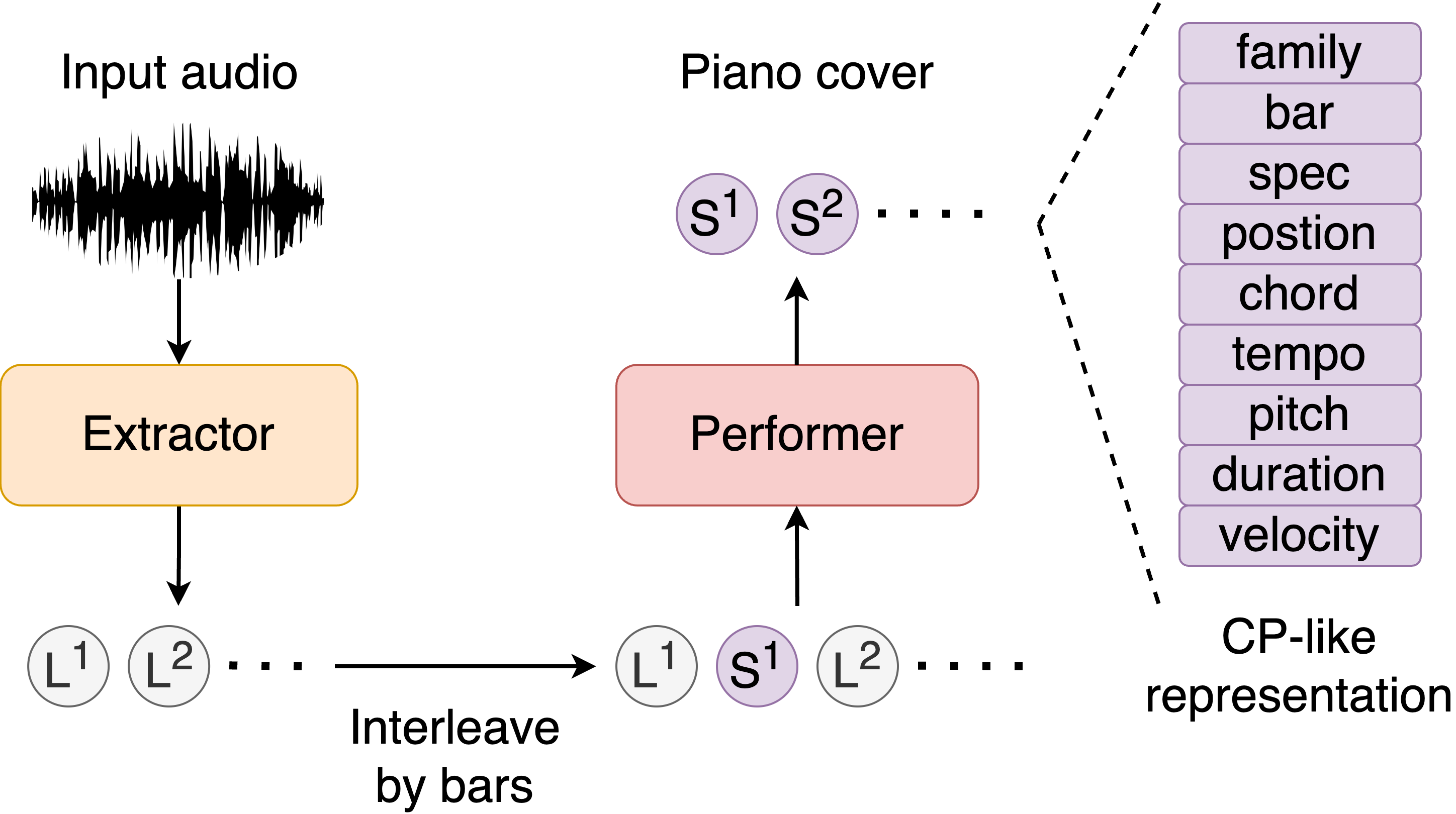}
  \caption{For each bar (musical measure) $k$ of the input, the Extractor transcribes from the input its lead sheet $L^k$ (a token sequence), and the Performer generates autoregressively the piano performance $S^k$ (also a token sequence) for the same bar given the current and preceding sequences of lead sheet $[L^1, L^2, \dots L^k]$ and the preceding piano performances $[S^1, S^2, \dots S^{k-1}]$ organized in an interleaving fashion.}
  \label{fig:repr}
\end{figure}

To better inform the Performer of the temporal correspondence between the lead sheet $L$ and the piano $S$, we follow the approach of Wu\,\&\,Yang~\cite{wu2023compembellish} and implement the Performer as a decoder-only Transformer that deals with an interleaved sequence composed of a ``bar-wise mix'' of $L$ and $S$, as shown in \figref{fig:repr}.
Specifically, we segment $L$ and $S$ respectively into multiple sub-sequences $L=[L^1, \dots, L^B]$ and $S=[S^1, \dots, S^B]$, where $L^k$ and $S^k$ are the lead sheet sub-sequence and piano sub-sequence for the same, $k$-th bar (i.e., musical measure) of the input music, and $B$ denotes the number of bars.
We use the bar-wise mix $[L^1, S^1, \dots, L^B, S^B]$ to train the Performer, such that the generation for the $k$-th bar of piano $S^k$ would depend on (i.e., can attend to) the current and preceding sub-sequences of lead sheet $[L^1, L^2, \dots L^k]$ and the preceding piano sub-sequences $[S^1, S^2, \dots S^{k-1}]$.


\subsection{Proposed Token Representation}

A known weakness of the \emph{MIDI-like} tokens \cite{huang2018music} used by Pop2Piano to represent piano music $S$ is the overly large number of tokens needed to represent a musical bar, 
making it hard to learn long-term dependency \cite{hsiao2021compound}.
We instead adopt a modified version of compound-word (CP) token representation 
\cite{hsiao2021compound} to make the token sequence  compact.
The main idea of CP is to group related tokens into a ``super token.'' 
As shown in \figref{fig:repr},
tokens belonging to the same super token are combined into a single embedding vector before feeding to the Transformer, which in turn predicts the tokens belonging to the next super token collectively at the same time using different heads.
We refer to \cite{hsiao2021compound} for details.

Specifically, CP uses the idea of ``token classes'' to organize the tokens~\cite{hsiao2021compound}. While the original CP considers only two token classes, metric and non-metric, 
we consider four token classes so as to have a unified CP-like representation for not only the piano $S$ but also the lead sheet $L$.
They are \tokencls{Spec}, \tokencls{Bar}, \tokencls{Metric} and \tokencls{Note}.
\tokencls{Spec} class includes special tokens, e.g., \token{BOS} and \token{EOS} (i.e., beginning or ending of a sequence).
\tokencls{Bar} class contains \token{bar\_src} and \token{bar\_tgt}, which are used to distinguish tokens belongs to lead sheet or piano cover.
\tokencls{Metric} class is composed with three sub-classes: \tokencls{Position}, \tokencls{Tempo}, and \tokencls{Chord}.
\tokencls{Note} class is composed with three sub-classes describing a musical note: \tokencls{Pitch}, \tokencls{Duration} and \tokencls{Velocity}.\footnote{\tokencls{Position} indicates the timing shift from the beginning of each bar, in the resolution of 16th note.
\tokencls{Tempo} indicate the beat-per-minute (BPM), 
ranging from 32 to 244.
\tokencls{Chord} is used to represent the chord condition, composed by \{chord root, chord quality\} pair. 
\tokencls{Pitch} ranges from A0 to C8.
\tokencls{Duration} indicates the length of each notes, with the same resolution of \tokencls{Position}.
We limit the maximum duration to the length of 8 quarter notes in this work.
\tokencls{Velocity} takes 
MIDI value from 0 to 127.
}

\subsection{Implementation Details}
To implement the Extractor of PiCoGen,
we employ SheetSage \cite{donahue2022melody}, the state-of-the-art for lead sheet transcription.
Authors of SheetSage have kindly released a model checkpoint\footnote{\url{https://github.com/chrisdonahue/sheetsage}} well-trained on the Hook Theory dataset~\cite{donahue2022melody}, which consists of approximately 40,000 paired data of audio clips and lead sheets.  We simply leverage this checkpoint as our Extractor without fine-tuning.


As for the Performer, we adopt the decoder-only architecture of the CP Transformer \cite{hsiao2021compound} and train it from scratch.
The model has 8 self-attention layers, 8 heads for multi-head attention, 512 hidden dimensions, a sequence length of 1,024 super tokens, and GeLU as the activation function.
At inference time, the lead sheet $L$ obtained from the Extractor is converted to the proposed CP-like tokens 
and then fed to the Performer to generate the piano cover $S$.

\begin{figure*}
  \centering
  \includegraphics[width=.98\linewidth]{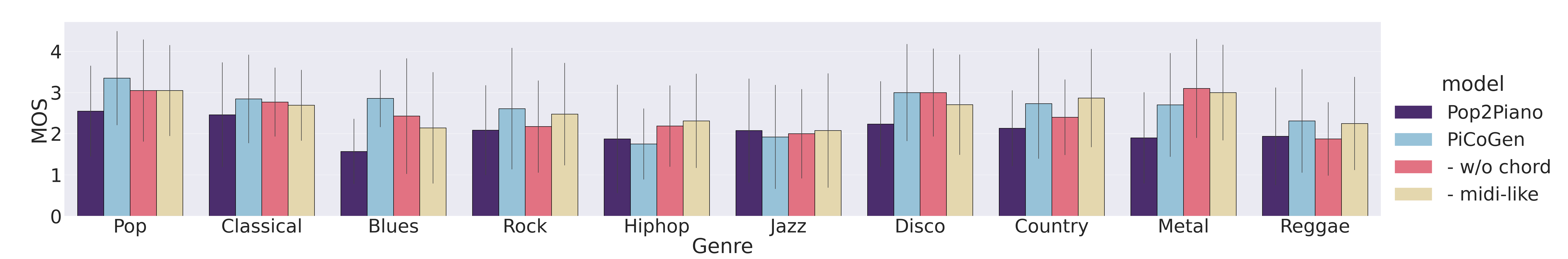}
  \vspace{-0.5mm}
  \caption{Mean opinion score in the metric OVL in the subjective evaluation, for each of the ten considered genres.}
  \label{fig:mos}
\end{figure*}

\section{Experimental Setup}

\subsection{Datasets}

Training the Performer requires paired data of lead sheets $\mathcal{D}_{l}$ and piano performances $\mathcal{D}_{p}$. For $\mathcal{D}_{p}$, we adopt the Pop1k7 dataset \cite{hsiao2021compound}, which encompasses around 1,700 piano covers of various Japanese anime, Korean popular and Western popular songs.
To construct $\mathcal{D}_{l}$, we follow the approach of Wu\,\&\,Yang~\cite{wu2023compembellish} and employ symbolic-domain music analysis techniques~\cite{uitdenbogerd1998manipulation} to recognize the lead sheet from each piano performance.

For model evaluation, we use the GTZAN dataset 
\cite{tzanetakis2002musical},\footnote{\url{http://marsyas.info/download/data\_sets}} a well-known public-domain dataset containing 100 30-second song clips for each of 10 selected genres, including Pop, Jazz, Classical, Rock, etc (see \figref{fig:mos} for the full list), enabling evaluation beyond Pop music.
Note that SheetSage assumes constant tempo and would report errors for songs that violate this assumption. We therefore discard the 117 songs of GTZAN that has this problem, using the remaining 883 songs in our evaluation. 


\subsection{Baseline and Ablations}

We consider Pop2Piano~\cite{choi2023pop2piano} as the baseline, as it represents the state-of-the-art for piano cover generation.
Specifically, we use the model checkpoint released by the original authors 
for testing.\footnote{\url{https://github.com/sweetcocoa/pop2piano}}
Moreover, we train and evaluate two ablated versions of PiCoGen.
The first version, denoted as ``w/o chord'', uses only the melody as the intermediate representation $L$, namely dropping chord information entirely. This is to study the effect of chords.
The second version, denoted as ``midi-like,'' uses the MIDI-like tokens~\cite{huang2018music} to represent the piano music, instead of the more advanced CP-like representation. This is to use the same token representation as Pop2Piano~\cite{choi2023pop2piano} so as to examine whether any possible performance difference between Pop2Piano and PiCoGen is due to the two-stage architecture or due to token representations. 
We note that these ablations involve changes only to the intermediate data, and that both ablations adhere to the proposed two-stage strategy. 

\subsection{Evaluation Metrics} \label{sec:objective}
For objective evaluation, we follow 
Pop2Piano and compute melody chroma accuracy (MCA)~\cite{choi2023pop2piano}. The  idea is to compare the top line of the piano MIDI generated by a model with the melody contour of the input audio. 
According to Pop2Piano~\cite{choi2023pop2piano}, the melody contour is computed by  separating the vocal  of the input via a source separation model~\cite{spleeter2020}, and then using
pYIN \cite{mauch2014pyin} for f0 estimation from the separated vocal. MCA is $\in [0,1]$ and the higher the better.

For subjective evaluation, 
we conduct an online user study involving 30 volunteers consisting of 10 amateurs, 17 pro-ams\cite{saivanka2023adoption}, and 3 pros.
We generate 31 distinct testing sets, distributing them to the human subjects according to their birth dates.
In each testing set, we randomly select 5 genres from the GTZAN dataset and designate one song per genre as the testing target. Consequently, each testing set comprises 5 songs, and we generate 4 results for each song, using Pop2Piano, PiCoGen and the two ablation models, presented to the subject in random order.
Subjects are instructed to score the results with a Likert scale from 1 to 5 (the higher the better) in the following five aspects:
\begin{itemize}
    \item \textbf{Similarity (SI)}: How much the piano cover sounds similar to its original song in general?
    \item \textbf{Melody similarity\,(SI$_\text{m}$)}: The perceived similarity between the melody lines of the piano cover and the original song.
    \item \textbf{Chord similarity\,(SI$_\text{c}$)}: The similarity in chord progression.
    \item \textbf{Music fluency (FL)}: Does the piano cover sound fluent?
    \item \textbf{Overall (OVL)}: How much do you like the piano cover?
\end{itemize}


\begin{table}[t]
    \centering 
    \begin{tabular}{ l|c|ccccc }
        \toprule
        \multirow{2}{*}{Model} & \emph{objective} & \multicolumn{5}{|c}{ \emph{subjective evaluation}\,$\uparrow$}  \\ 
        & MCA\,$\uparrow$ & SI & SI$_\text{m}$ & SI$_\text{c}$ & FL & OVL \\
        \midrule
        Pop2Piano\cite{choi2023pop2piano} & \textbf{0.25} & 2.65 & 2.85 & 2.65 & 2.60 & 2.55 \\
        \midrule
        PiCoGen & 0.17 & \textbf{3.20} & \textbf{3.35} & \textbf{3.05} & \textbf{3.20} & \textbf{3.35}  \\
        \,\,\,\,--\textit{w/o chord} & 0.14 & 3.10  & 3.25 & 2.95 & 3.10 & 3.05 \\
        \,\,\,\,--\textit{midi-like} & 0.12 & 2.95 & 3.10 & 2.95 & 3.10 & 3.05 \\
        \bottomrule
    \end{tabular}
    \caption{Evaluation result on piano cover generation for only Pop music. Best result per metric highlighted in bold.}
    \label{table:pop}
\end{table}

\section{Results and Discussion}

\tabref{table:pop} displays the average MCA and mean opinion scores (MOS) from the user study for Pop only, the genre targeted by Pop2Piano.
We see that the proposed PiCoGen and its variants show a lower MCA than Pop2Piano, but outperform it greatly in every subjective metric.
Although statistical test does not reveal significant performance difference (due to large stanadard deviation), we view this as an encouraging result since PiCoGen does not use paired data of music and their piano covers at all.
Moreover, although the two ablated variants receive moderately lower MOS than PiCoGen, the result is in general comparable, providing further evidences the effectiveness of the two-stage approach.
We also note that the MOS in OVL reduces from 3.35 to 3.05 when dropping chords from the intermediate representation, suggesting the superiority of using lead sheet instead of melody only in PiCoGen.


\tabref{table:pop} also suggests a conflicting result between~MCA~and~its subjective counterpart SI$_\text{m}$.
PiCoGen underperforms Pop2Piano in MCA, but its MOS in SI$_\text{m}$ is higher.
We conjecture that this might due to the fact that MCA only considers vocal melody, neglecting the melody of the leading instrument, as discussed in \cite{pop909-ismir2020}.
We consider the development of better objective metrics as a future work and focus on the result of the user study here.



\figref{fig:mos} shows the MOS in OVL for all the ten genres in GTZAN, extending our evaluation beyond Pop music.
\figref{fig:mos} shows that PiCoGen outperforms Pop2Piano not only for Pop but also for genres such as Blues, Disco, and Metal, demonstrating the generalizability of PiCoGen to some extent. 
However, there are also many genres for which Pop2Piano and PiCoGen have similar MOS.
In particular, both Pop2Piano and PiCoGen perform poorly for Hip-hop music.
Upon scrutinizing the attributes of the testing songs, we identify that the quality of the lead sheets extracted by SheetSage is not satisfactory for some of the genres, especially for Hip-hop.
Moreover, 
the MOS of PiCoGen in \tabref{table:pop} and \figref{fig:mos} ranges from 2.00 to 3.50 in general, leaving considerable room for future improvement given that the scores are from 1 to 5. 

There are many interesting avenues to be investigated in future work.
First, both Pop2Piano and PiCoGen can be potentially improved by collecting a larger training dataset encompassing more genres.
However, we note that Pop2Piano demands both the piano covers and their original songs in diverse genres as the paired data for training. 
In contrast, it is easier to expand the training data of PiCoGen as we only need the piano covers (not the audio recordings of the original songs).
Second, besides data, better result for PiCoGen might be obtained along with advances in lead sheet transcription, and by incorporating intermediate representations of music other than the lead sheet (e.g., rhythmic elements) for musical genres with no dominant melody line.
Finally, it might also be interesting to develop a model that combines the strengths of the music-informed design of PiCoGen and the data-driven design of Pop2Piano, leveraging both unpaired and paired data for training.



\section{Conclusion}
In this paper, we have presented PiCoGen, a  generative model capable of creating a piano cover of an input audio with two distinct steps: extracting a lead sheet from the audio and generating a piano performance based on this lead sheet.
Treating the lead sheet as the common ground between the input music 
and the target piano MIDI, 
PiCoGen bypasses the need of curating paired data of  covers and their original songs for training.
Compared to the existing single-stage model Pop2Piano, PiCoGen exhibits comparable or  superior performance for Pop music and some other  genres.
Directions for future improvement have also been discussed.



\bibliographystyle{ACM-Reference-Format}
\balance
\bibliography{icmr2024}










\end{document}